\documentstyle[aps,preprint]{revtex} %% RevTeX V 3.1

\font\msbm=msbm10 scaled 1200 \font\eusb=eusb10 scaled 1200
\def\KA{\kappa^2} \def\Mi#1{{\cal M}^{#1}} \def\Ns#1{\Z{\text{#1}}}
\def\Z#1{_{\lower2pt\hbox{$\scr#1$}}} \def\fnu{{\mu}} \def\Vf{\left(2V\right)}
\def\ns#1{_{\text{#1}}} \def\az{\case12a^2} \def\Az{(\az)} \def\zb{x}
\def\Der#1#2{{#1\hphantom{#2}\over#1#2}}
\def\DDer#1#2#3{{#1^{#3}\hphantom{#2}\over#1#2^{#3}}} 
\def\NN{{\cal N}} \def\Vv{{\cal V}} \def\dd{{\rm d}} \def\e{{\rm e}}
\def\rarr{\rightarrow} \def\GA{\Gamma}  \def\OO{{\rm O}}
\def\goesas{\mathop{\sim}\limits} 
\def\PNB{\Psi\Ns{NB}} \def\PTL{\Psi\Ns{TL}} \def\PTV{\Psi\Ns{TV}}
\def\dsp{\displaystyle} \def\scr{\scriptstyle} \def\be{\beta}
\def\PS{\Psi} \def\pt{\partial} \def\ph{\phi} \def\si{\sigma} \def\rh{\rho}
\def\Ss#1#2#3{\hbox{\eusb S}^{#1}_{#2,#3}} \def\Mfn#1{M\!\left(#1\right)}
\def\Cc#1#2#3{\hbox{\eusb C}^{#1}_{#2,#3}} \def\Ufn#1{U\!\left(#1\right)}
\def\el{\ell}\def\Dn{{\cal D}_{n,i}}\def\al{\alpha} \def\Aa{{\cal A}}
\def\Zop{\hbox{\msbm Z}} \def\w#1{\;\hbox{#1}\;} \def\Hh{{h}}
\def\Pp{{\cal P}}\def\Pn{\Pp_{n,i}} \def\Uu{{\cal U}}\def\UU{\Uu(a,\ph)}
\def\asym{\left[{(-\Uu)^{3/2}\over3V}-{\pi\over4}\right]}
\def\gsim{\mathop{\hbox{${\lower3.8pt\hbox{$>$}}\atop{\raise0.2pt\hbox{$
\sim$}}$}}} \def\PR#1{Phys.\ Rev.\ {\bf#1}}\def\NP#1{Nucl.\ Phys.\ {\bf#1}}
\def\PL#1{Phys.\ Lett.\ {\bf#1}} \def\AP#1#2{Ann.\ Phys.\ (#1) {\bf#2}}
\def\NC#1{Nuovo Cimento {\bf#1}}\def\CQG#1{Class.\ Quantum Grav.\ {\bf#1}}

\begin{document}
\title{Wave functions for arbitrary operator ordering in the de Sitter
minisuperspace approximation$^{\dag}$}
\author{David L. Wiltshire$^*$}
\address{Department of Physics and Mathematical Physics, University of
Adelaide, Adelaide S.A. 5005, Australia.}

\maketitle \hyphenation{}

\begin{abstract}
\baselineskip=14pt plus.2pt minus.2pt

We derive exact series solutions for the Wheeler-DeWitt equation corresponding
to a spatially closed Friedmann-Robertson-Walker universe with cosmological
constant for arbitrary operator ordering of the scale factor of the universe.
The resulting wave functions are those relevant to the approximation which has
been widely used in two-dimensional minisuperspace models with an inflationary
scalar field for the purpose of predicting the period of inflation which
results from competing boundary condition proposals for the wave function of
the universe. The problem that Vilenkin's tunneling wave function is not
normalizable for general operator orderings, is shown to persist for other
values of the spatial curvature, and when additional matter degrees of freedom
such as radiation are included.
\vskip2\baselineskip
\leftline{$\dag$ Based on a talk presented at the
{\it 2nd Australasian Conference on General Relativity and}}
\leftline{{\it Gravitation}, Sydney, July 1998.}
\leftline{To appear with the Proceedings in {\it General Relativity and
Gravitation}}\medskip
\leftline{* Electronic address: dlw@physics.adelaide.edu.au}
\vskip\baselineskip\hrule

\end{abstract}

\global\preprintstyfalse\baselineskip=16pt plus.2pt minus.2pt

\section{Introduction}

Over the past year new developments concerning the possibility of open
inflation \cite{HT1,Lin1} in quantum cosmology have led to renewed interest
in the debate\cite{Lin1,HT2,discord} about the relative merits of competing
proposals for boundary conditions on the wave function of the Universe. This
debate has centred on the differing predictions of the ``no boundary'' proposal
of Hartle and Hawking \cite{HH}, and the ``tunneling'' proposals of Vilenkin
\cite{Vil33,Vil37} and of Linde \cite{Lin2}, which are themselves distinct.

Much of the debate between proponents of the competing proposals arises from
differing predictions obtained using probability amplitudes which are assumed
to apply to the respective wave functions. In particular, the nucleation
probability for instanton-dominated transitions which describe a universe
which ``tunnels from nothing'' is assumed to be
\begin{equation}
\hbox{\eusb P}\propto|\PS|^2\propto\cases{\e^{-2I\ns{cl}}&$\PNB$\cr
\e^{+2I\ns{cl}}&$\PTL,\ \PTV$\cr}\label{Iproba}
\end{equation}
where the subscripts (NB), (TL) and (TV) refer to the no-boundary wave function
and the tunneling wave functions of Linde and Vilenkin respectively. For the
solutions in question, which correspond to a model in which gravity is coupled
to a scalar field, $\ph$, with potential, $V(\ph)$ in dimensionless units (see
(\ref{action}) below for our conventions), the Euclidean action of the
instanton is
\begin{equation}
I\ns{cl}={-1\over3V(\ph\Z0)}\,,\label{Iprobb}
\end{equation}
$\ph\Z0$ being the value of the scalar field at nucleation.

It has been pointed out by supporters of Vilenkin's proposal
\cite{discord,bh2} that arguments which use (\ref{Iproba}) as their starting
point are sometimes ill-founded, since the identifications assumed in
(\ref{Iproba}) are not supported by more detailed minisuperspace calculations
in models such as that of black hole nucleation in the very early universe
\cite{bh1,bh2}. In particular, it appears that $\PTV$ cannot be identified with
the probability amplitude (\ref{Iproba}) in this case, thereby side-stepping
the criticism of \cite{bh2}. We have pointed out recently \cite{KW}, however,
that similar criticisms apply to the model involving the nucleation of a closed
Friedmann-Robertson-Walker universe with an inflationary scalar field --
the ``work horse model'' of numerous discussions in quantum cosmology.

The conclusion drawn in \cite{KW} was that the identification of $\PTV$,
(or indeed of $\PTL$), with the probability amplitude (\ref{Iproba}) depends
on Planck scale physics through the choice of operator ordering in the
minisuperspace Wheeler-DeWitt equation. This would not be a serious problem
if there was a natural choice of operator ordering for which the identification
(\ref{Iproba}) could still be made in the case of $\PTV$. However, the choice
of ordering which to our knowledge is the only one to have been claimed as
natural in this minisuperspace model \cite{HP1,Louko,Bar}, is one for which
$\PTV$ cannot be normalized, the choice of normalization being central to the
identification (\ref{Iproba}). The grounds for ``naturalness'' of the relevant
factor ordering include 1-loop unitarity \cite{Bar}. In fact, the same operator
ordering was assumed in calculations which claim to lead to the suppression of
values of the scalar field with a potential, $V(\ph)$, above the Planck scale
in the relevant probability amplitude \cite{BK1} -- or even to lead to an
enhancement of sub-Planckian values of $V(\ph)$ in a conformally coupled model
\cite{BK2} -- arguments which have usually been assumed \cite{Vil37,BK1,BK2} to
favour Vilenkin's tunneling wave function, $\PTV$. Our observation \cite{KW},
which concerns Planck scale corrections in the scale factor, $a$, rather than
in the scalar field $\ph$ (as assumed in \cite{Bar,BK1,BK2}), casts
considerable doubt on such claims. Thus the ``prediction of inflation'' cannot
be claimed as a success of Vilenkin's wave function, as is often widely
assumed.

The conclusions of ref.\ \cite{KW} were based on an analysis of the second
order ordinary differential equation obtained in the ``de Sitter minisuperspace
approximation'' for arbitrary ordering of the scale factor in the
Wheeler-DeWitt equation. Our analysis was based on the qualitative behaviour
of this equation in various relevant limits. In fact, exact series solutions
may be developed for the models of interest. It is our intention to present
these solutions in this paper, and to relate the solutions to the results of
\cite{KW}.

\section{Minisuperspace model}

The specific model considered here is the 2-dimensional minisuperspace
corresponding to the classical action for gravity coupled to a scalar field,
\begin{equation}
S={1\over4\KA}\left[\;\int\limits_{\Mi{}}\dd^4x\sqrt{-g}
{\cal R}+2\int\limits_{\pt\Mi{}}\dd^3x\sqrt{h}\,{\cal K}\right]
+{3\over\KA}\int\limits_{\Mi{}}\dd^4x\sqrt{-g}\left(
-{1\over2}g^{\mu\nu}\pt_\mu\ph\pt_\nu\ph-{V(\ph)\over2\si^2}\right),
\label{action}\end{equation}
where $\KA=4\pi G=4\pi m^{-2}\ns{Planck}$, $\cal K$ is the trace
of the extrinsic curvature, and the metric is assumed to take the closed
Friedmann-Robertson-Walker form
\begin{equation}
\dd s^2=\si^2\left\{-\NN^2\dd t^2+a^2(t)\dd{\Omega\Z3}^2\right\}\,,\label{FRW}
\end{equation}
where $\dd{\Omega\Z3}^2$ is a round metric on the 3-sphere, and $\si^2=\KA/(6
\pi^2)$.

The Hamiltonian constraint obtained from the $(3+1)$-decomposition of the
field equations may be quantized to yield the Wheeler-DeWitt equation
\begin{equation}
\left[{1\over a^p}\Der\pt a\, a^p\Der\pt a-{1\over a^2}\DDer\pt\ph2
-a^2\UU\right]\PS=0,\label{WdW}
\end{equation}
where
\begin{equation}
\UU\equiv 1-a^2V(\ph), \label{pot}
\end{equation}
and we have allowed for possible operator-ordering ambiguities through the
integer power, $p$, in the first term. The approximation that has been adopted
in discussions such as that of the prediction of inflation from quantum
cosmology \cite{Vil33,Vil37} is to confine the discussion to
regions in which the the potential $V(\ph)$ can be approximated by a
cosmological
constant, so that the $\ph$ dependence in (\ref{WdW}) can be effectively
ignored and a standard 1-dimensional WKB analysis applied.

The Wheeler-DeWitt equation for the ``de Sitter minisusperspace approximation''
has been solved exactly in terms of Airy functions by Vilenkin \cite{Vil37}
for the special case of operator ordering $p=-1$. We wish to observe here that
similar series solutions can be readily obtained for the case of arbitrary
factor ordering. In particular, with the redefinition
\begin{equation}
\PS\equiv z^{-(p-1)/4}y(z),\label{ydef}
\end{equation}
where $z\equiv\az$, we find that the 1-dimensional Wheeler-DeWitt resulting
from (\ref{WdW}) when the $\phi$--dependence is neglected is given by
\begin{equation}
z^2{\dd^2y\over\dd z^2}+z{\dd y\over\dd z}+\left(2Vz^3-z^2-\nu^2\right)y=0,
\label{Bocher25}\end{equation}
where $\nu=|\case14(p-1)|$. This equation is classified as a type $\{2 5\}$
B\^ocher equation in the terminology of Moon and Spencer \cite{MS}, although
its solution has not been explicitly tabulated. Near the regular singular
point at $z=0$, the solution $y(z)$ approaches that of the modified Bessel
equation, since for finite $V$ the term involving $V$ in (\ref{Bocher25}) is
sub-dominant.

\section{Convergent series solutions}

It is straightforward to apply the Frobenius method to (\ref{Bocher25}) to
obtain series solutions, although the calculations are rather lengthy. In
particular, after a considerable amount of algebra we find that the two
linearly independent solutions may be constructed in terms of the power series
$y_+(z)$ and $y_-(z)$ given by
\begin{equation}
y_\pm(z)=z^{\pm\nu}\left\{1+\sum_{\el=1}^\infty\left[\Aa^\pm_{2\el}z^{2\el}-
\Aa^\pm_{2\el+1}z^{2\el+1}\right]\right\} \label{ypmdef}
\end{equation}
where the coefficients of the even and odd powers of $z$ are respectively
\begin{equation}
\Aa^\pm_{2\el}=\sum_{i=0}^{[\el/3]}\Vf^{2i}\Ss{2i}\pm\el,
\end{equation}
\begin{equation}
\Aa^\pm_{2\el+1}=\sum_{i=0}^{[(\el-1)/3]}\Vf^{2i+1}\Ss{2i+1}\pm\el,
\end{equation}
$[\quad]$ denotes the integer part, and the sums $\Ss j\pm\el$ are
defined iteratively by
\begin{eqnarray}
\Ss0\pm\el&=&\Cc0\pm{2\el},\qquad\qquad \el\ge0,\\
\Ss{2i+1}\pm\el&=&\sum_{j=3i}^{\el-1}\Cc{2j+1}\pm{2\el+1}\Ss{2i}\pm j,
\qquad\qquad\el\ge1,\quad0\le i\le[(\el-1)/3],\\
\Ss{2i}\pm\el&=&\sum_{j=3i-1}^{\el-1}\Cc{2j}\pm{2\el}\Ss{2i-1}\pm{j-1}
\qquad\qquad\el\ge3,\quad0\le i\le[\el/3],\label{Slast}\\
\end{eqnarray}
in terms of the quantities $\Cc j\pm\el$, defined by
\begin{equation}
\Cc j+k={j!!(j+2\nu)!!\over k!!(k+2\nu)!!}\equiv{\GA\left(\case j2+1
\right)\GA\left(\case j2+\nu+1\right)\over2^{k-j}\GA\left(\case k2+1\right)
\GA\left(\case k2+\nu+1\right)},\qquad\qquad \nu\ge0
\end{equation}
and
\begin{equation}
\Cc j-k={j!!(j-2\nu)!!\over k!!(k-2\nu)!!}\equiv{\GA\left(\case j2+1
\right)\GA\left(\case j2-\nu+1\right)\over2^{k-j}\GA\left(\case k2+1\right)
\GA\left(\case k2-\nu+1\right)},\qquad\qquad \nu>0,\quad\nu\ne1,\case32,2,
\case52,\dots\label{Cm}
\end{equation}
$j$ and $k$ being two positive integers which differ by an even number.
For the special cases $\nu=1,\case32,2,\case52,\dots$, the quantities
(\ref{Cm}) must be replaced by
\begin{equation}
\Cc j-k=\cases{\hbox{$\dsp {(-1)^{(k-j)/2}\GA\left(\case j2+1\right)
\GA\left(\nu-\case k2\right)\over2^{k-j}\GA\left(\case k2+1\right)
\GA\left(\nu-\case j2\right)}$},&if $k<2\nu$; or $k>2\nu$ and $2\nu-k$ is odd,
\cr 0,&if $k=2\nu$; or $k>2\nu$, $j<2\nu$ and $2\nu-k$ is even,\cr
\hbox{$\dsp{\GA\left(\case j2+1\right)\GA\left(\case j2-\nu+1\right)\over2^{k-j}
\GA\left(\case k2+1\right)\GA\left(\case k2-\nu+1\right)}$},
&if $k>2\nu$, $j\ge2\nu$ and $2\nu-k$ is even,\cr}
\end{equation}

The linearly independent solutions to (\ref{Bocher25}) are then
\begin{equation}
Y\Z1(z)=y_+(z) \label{Ya}
\end{equation}
and
\begin{equation}
Y\Z2(z)=\cases{y_-(z),&$\nu\ne0,1,\case32,2,\case52,\dots$\cr
y_*(z),&$\nu=0$\cr
y_-(z)+{\cal B}_-y_*(z),&$\nu=1,\case32,2,\case52,\dots$\cr} \label{Yb}
\end{equation}
where
\begin{equation}
{\cal B}_-={1\over2\nu}\left(\Aa^-_{2\nu-2}-2V\Aa^-_{2\nu-3}\right),\qquad
\nu\ge1
\end{equation}
(with $\Aa^-_0=1$, $\Aa^-_{\pm1}=0$), and the series $y_*(z)$ is defined by
\begin{equation}
y_*(z)=y_+(z)\ln(z)-2\sum_{n=2}^\infty\sum_{i=1}^{\Hh_{n}}
{(-2V)^{n-2\Dn}\over\prod_{j=1}^{\Dn}\al^i_j\left(\al^i_j
+2\nu\right)}\left(\sum_{k=1}^{\Dn}{\left(\al^i_k+\nu\right)\over\al^i_k
\left(\al^i_k+2\nu\right)}\right)z^{n+\nu}.\label{ysing}
\end{equation}
The quantities in the summation in the coefficient of $z^{n+\nu}$ in
(\ref{ysing}) are defined in terms of the sequences of all partial sums
obtained from partitioning the integer $n$ by $2$ or $3$. In particular, let
\begin{equation}
\Pn=\left\{\al^i_j \in \Zop\ |\ \al^i_1<\al^i_2<\dots<\al^i\Z{\Dn}=n,\ \al^i_1
\in\{2,3\},\ (\al^i_{j+1}-\al^i_j)\in\{2,3\}\right\}
\end{equation}
where $\Dn=\w{dim}\Pn$. The index $i=1,2,\dots,\Hh_n$ labels the different
possible distinct ordered sequences of such integers, the number of which is
given by
\begin{equation}
\Hh_n=\sum_{\hbox{$\scr r=0,\ s=0$}\atop\hbox{$\scr3r+2s=n$}}^{[n/3],\ [n/2]}
{(r+s)!\over r!s!}\,,
\end{equation}
i.e., the $n$th element of the Padovan sequence \cite{Gr,St}, $\{1,0,1,1,1,2,2,
3,\dots\}$, defined recursively by $\Hh_0=\Hh_2=1$, $\Hh_1=0$, $\Hh_n=\Hh_{n-3}
+\Hh_{n-2}$ for $n\ge3$. The sequences in $\Pp_n$ may be constructed
iteratively by simply appending the integer $\{n\}$ to each sequence in
$\Pp_{n-3}$ and in $\Pp_{n-2}$, as shown in Table 1.

For $p=1$, (i.e., $\nu=0$), for example, we find solutions given by (\ref{Ya})
and (\ref{Yb}) with
\begin{eqnarray}
y_+(z)= &1&+\case{1}{4}z^2-\case{1}{9}2Vz^3+\case{1}{64}z^4-\case{13}{900}2Vz^5
+\left[\case{1}{2304}+\case{1}{324}\Vf^2\right]z^6-\case{433}{705600}2Vz^7
\nonumber\\ &&+\left[\case{1}{147456}+\case{71}{259200}\Vf^2
\right]z^8-\left[\case{2957}{228614400}+\case{1}{26244}\Vf^2\right]2Vz^9
\nonumber\\ &&+\left[\case{1}{14745600}+\case{11273}{1270080000}\right]
\Vf^2z^{10}+\dots\\
y_*(z)=&y_+&(z)\ln(z)-\case{1}{4}z^2+\case{2}{27}2Vz^3-\case{3}{128}z^4
+\case{253}{13500}2Vz^5-\left[\case{11}{13824}+\case{1}{324}\Vf^2\right]z^6
\nonumber\\ &&+\case{153527}{148176000}2Vz^7-\left[\case{25}{1769472}+
\case{2123}{5184000}\Vf^2\right]z^8+\left[\case{3671179}{144027072000}+
\case{11}{236196}\Vf^2\right]2Vz^9\nonumber\\ &&-\left[\case{137}{884736000}
+\case{2886157}{177811200000}\Vf^2\right]z^{10}+\dots
\end{eqnarray}

\begin{table}
\caption{Integer sequences contributing to $y_*$} \begin{tabular}{lll}
$n$&$\Hh_n$&$\Pn$, $i=1,\dots,\Hh_n$\\ \tableline
$2$&$1$&$\{2\}$\\
$3$&$1$&$\{3\}$\\
$4$&$1$&$\{2,4\}$\\
$5$&$2$&$\{2,5\}$, $\{3,5\}$\\
$6$&$2$&$\{3,6\}$, $\{2,4,6\}$\\
$7$&$3$&$\{2,4,7\}$, $\{2,5,7\}$, $\{3,5,7\}$\\
$8$&$4$&$\{2,5,8\}$, $\{3,5,8\}$, $\{3,6,8\}$, $\{2,4,6,8\}$\\
$9$&$5$&$\{3,6,9\}$, $\{2,4,6,9\}$, $\{2,4,7,9\}$, $\{2,5,7,9\}$, $\{3,5,7,9\}$
\\
$10$&$7$&$\{2,4,7,10\}$, $\{2,5,7,10\}$, $\{3,5,7,10\}$, $\{2,5,8,10\}$,
$\{3,5,8,10\}$, $\{3,6,8,10\}$, $\{2,4,6,8,10\}$\\
\vdots&\vdots&\vdots\\
\end{tabular} \end{table}

\section{Asymptotic series and the WKB limit}

Although the series solutions of Sec.\ III converge for all positive values
of $a$, the rate of convergence is extremely slow and for many purposes a
knowledge of the asymptotic series is desirable. For example, boundary
conditions are often formulated in terms of the WKB solutions, which are
effectively the leading order terms in such limits. Since the nature of the
WKB solutions varies according to the sign of $\UU$ (\ref{pot}), instead of
using the variable $z$ of Sec.\ III, it is convenient to introduce a new
variable
\begin{equation}
\zb\equiv\Vf^{-2/3}\left(1-2Vz\right),
\end{equation}
and to make the transformation $y\equiv z^{-1/2}u$. We then see that
(\ref{Bocher25}) may be written equivalently as
\begin{equation}
{\dd^2u\over\dd\zb^2}-\left[\zb+{\Vf^{4/3}\fnu\over\left[\Vf^{2/3}\zb-1\right]
^2}\right]u=0
\label{Airy25}\end{equation}
where
\begin{equation}
\fnu\equiv\nu^2-\case14=\case1{16}(p+1)(p-3).
\end{equation}
The essential point to note is of course that to leading order the asymptotic
series derived from (\ref{Airy25}) as $\zb\rarr\pm\infty$ are no different to
those of the special case $\fnu=0$, or $p=-1,3$, for which $u(\zb)$ is an
exact Airy function \cite{Vil37}.

The first limit of interest is $\zb\rarr\infty$, which, when physically
relevant, corresponds to large finite values of $a$ such that $a^2V\ll1$
nonetheless. This requires us to take $V\ll1$, i.e., to restrict the scalar
field, $\ph$, to values with a potential much below the Planck scale, which
is of course physically justified.

Since $\PS\equiv\left(\az\right)^{-(p+1)/4}u(\zb)$, we find that the
appropriate asymptotic series for $\PS$ are linear combinations of
the two modes
\begin{eqnarray}
\PS_\pm\propto{\exp\left[{\pm\Uu^{3/2}\over3V}\right]\over
a^{(p+1)/2}\Uu^{1/4}}\Bigl\{1\pm&&\left(\case5{48}-\case13\fnu\right){2V\over
\Uu^{3/2}}\mp\case25\fnu{2V\over\Uu^{5/2}}\nonumber\\ &&+\left(\case{77}{96}-
\case16\fnu\right)\left(\case5{48}-\case13\fnu\right){\Vf^2\over\Uu^3}\mp
\case37\fnu{2V\over\Uu^{7/2}}+\dots\Bigr\}
\label{AiryE}\end{eqnarray}
The leading order term is recognized as the WKB mode in the tunneling region.
In practice, the requirement that $a^2V\ll1$ means that this expansion is of
little relevance beyond the leading term: terms of order $V$ in the numerator
occur for arbitrarily large odd half-integral powers of $\Uu$ in the
denominator, and similarly for terms of order $V^2$.

The question as to which linear combination of the modes (\ref{AiryE})
correspond to the respective solutions $\PS\Z1=a^{(p-1)/4}Y\Z1\Az$
and $\PS\Z2=a^{(p-1)/4}Y\Z2\Az$ of (\ref{Bocher25}) can be
resolved by expanding them in powers of $2Vz\equiv a^2V$. In particular,
from (\ref{ypmdef})--(\ref{ysing}) we find $Y\Z1(z)$ and $Y\Z2(z)$ are
given by their definitions (\ref{Ya}) and (\ref{Yb}) where now
\begin{eqnarray}
y_{\pm}(z)&=&z^{\pm\nu}\sum_{n=0}^\infty\left(-2Vz\right)^n\sum_{\el=n}^\infty
\Ss n\pm{\el+[n/2]}z^{2\el},\nonumber\\
&=& I_{\pm\nu}(z)-2Vz\sum_{\el=1}^\infty\sum_{j=0}^{\el-1}\Cc 1\pm{2\el+1}
\Cc0\pm{2j}z^{2\el}+\dots
\end{eqnarray}
and
\begin{eqnarray}
y_*(z)=&&y_+(z)\ln(z)\nonumber\\
&&-2z^{\nu}\sum_{n=0}^\infty(-2Vz)^n\sum_{\el=1+[n/2]}^\infty
\sum_{\hbox{$\scr i=1,$}\atop\hbox{$\scr{\cal D}_{2\el+n,i}=\el$}}^{\Hh_{2\el+
n}}{z^{2\el}\over\prod_{j=1}^{\el}\al^i_j\left(\al^i_j+2\nu\right)}\left(\sum_
{k=1}^{\el}{\left(\al^i_k+\nu\right)\over\al^i_k\left(\al^i_k+2\nu\right)}
\right).
\end{eqnarray}
We use the fact that 
\begin{equation}
I_{\pm\nu}\Az={1\over\sqrt{\pi}\,a}\exp\Az\left[1+
\OO(a^{-2})\right]
\end{equation}
and
\begin{equation}
K_{\nu}\Az={\sqrt{\pi}\over a}\exp(-\az)\left[1+\OO(a^{-2})\right]
\end{equation}
for large $a$ to
compare the leading terms with (\ref{AiryE}), noting that the overall
exponential $\exp\left[\pm\Uu^{3/2}\over3V\right]$ in (\ref{AiryE}) may be
expanded as $\exp\left[{\pm1\over 3V}\mp\az\left(1+\OO(a^2V)\right)
\right]$. In this way we see, similarly to the observation in \cite{KW}, that
the leading order terms agree if we identify
\begin{equation}
\PS_-\propto{1\over a^{(p-1)/2}}\left[Y\Z1\Az+\al Y\Z2\Az\right],
\label{BesselE1}\end{equation}
$\al\ne-1$, and
\begin{equation}
\PS_+\propto{1\over a^{(p-1)/2}}\left[Y\Z2\Az-Y\Z1\Az\right].
\label{BesselE2}\end{equation}
An overall multiplicative factor $\exp\left(\mp1\over 3V\right)$ must be
included in the constant of proportionality in (\ref{AiryE}) to cancel the
overall $\ph$ dependence in the cases that the wave function can be normalized.

In the limit $\zb\rarr-\infty$, which corresponds to $a^2V\gg1$, or $a\rarr
\infty$, we may find asymptotic series which apply beyond leading order. In
particular, the linearly independent modes are found to be
\begin{eqnarray}
\PS_-&\propto&{2\over a^{(p+1)/2}(-\Uu)^{1/4}}\left\{\cos\asym A\Z1+
\sin\asym A\Z2\right\},\label{AiryL1}\\
\PS_+&\propto&{1\over a^{(p+1)/2}(-\Uu)^{1/4}}\left\{-\sin\asym A\Z1+
\cos\asym A\Z2\right\},\label{AiryL2}
\end{eqnarray}
where
\begin{eqnarray}
A\Z1\equiv1&-&\left(\case{77}{96}-\case16\fnu\right)\left(\case5{48}-\case13
\fnu\right){\Vf^2\over(-\Uu)^3}+\case18\left(\case{16}{15}\fnu-\case{13}3
\right)\fnu{\Vf^2\over(-\Uu)^4}+\case1{10}\left(\case{78}{35}\fnu-\case{445}
{56}\right)\fnu{\Vf^2\over(-\Uu)^5}\nonumber\\&+&\left(\case{437}{192}-\case1
{12}\fnu\right)\left(\case{221}{144}-\case19\fnu\right)\left(\case{77}{96}-
\case16\fnu\right)\left(\case5{48}-\case13\fnu\right){\Vf^4\over(-\Uu)^6}+
\case1{12}\left(\case{1208}{315}\fnu-\case{113}{9}\right)\fnu{\Vf^2\over(-\Uu)
^6}\nonumber\\ &+&\dots,\label{Asum1}\\
A\Z2\equiv\phantom{1}&\phantom{-}&
\left(\case5{48}-\case13\fnu\right){2V\over(-\Uu)^{3/2}}+\case25
\fnu{2V\over(-\Uu)^{5/2}}-\case37\fnu{2V\over(-\Uu)^{7/2}}+\case49\fnu{2V\over
(-\Uu)^{9/2}}\nonumber\\&-&\left(\case{221}{144}-\case19\fnu\right)\left(\case
{77}{96}-\case16\fnu\right)\left(\case5{48}-\case13\fnu\right){\Vf^3\over(-
\Uu)^{9/2}}-\case5{11}\fnu{2V\over(-\Uu)^{11/2}}\nonumber\\ &+&\case1{99}\left(
\case{11}{5}\fnu^2-\case{1471}{40}\fnu+\case{28231}{256}\right)\fnu{\Vf^3\over
(-\Uu)^{11/2}}+\dots,\label{Asum2}
\end{eqnarray}
The phase in (\ref{AiryL1}) and (\ref{AiryL2}) is obtained by analytic
continuation of the corresponding modes in (\ref{AiryE}), taking care with
branch cuts similarly to the $\mu=0$ pure Airy function case \cite{Di}. To
leading order, this of course agrees with the results obtained using the WKB
connection formulae \cite{KW}. The no boundary wave function, $\PNB$, of Hartle
and Hawking corresponds to $\PS_-$, whereas Vilenkin's tunneling wave function,
$\PTV$, corresponds to $\case12\PS_-+i\PS_+$, for the modes given by
(\ref{AiryL1})--(\ref{Asum2}).

\section{Discussion}

The physical interpretation of the series solutions to the de Sitter
minisuperspace Wheeler-DeWitt equation (\ref{ydef}), (\ref{Bocher25}) yield no
surprises in comparison with the conclusions which have already been reached
in \cite{KW}. In particular, for operator orderings with $p\ge1$ the
contribution from $Y\Z2\Az$ diverges at the origin, causing $\PS$ to
diverge also. (For $p<1$ the divergence is regulated by the prefactor.) Thus
any boundary condition proposal which requires regularity of the wave function
as $a\rarr0$ must reduce to (\ref{BesselE1}) with $\al=0$ if $p\ge1$ in this
particular model; i.e., it must coincide with the proposal of Hartle and
Hawking to the level of the approximations which have been assumed. Since
normalizability of the wave function at $a=0$ is an essential ingredient of
calculations of the probability of inflation from quantum cosmology, the
question of operator ordering cannot be ignored in such calculations. If
the factor ordering $p=1$ is the natural one in the present model, as has been
claimed \cite{Louko,Bar}, then there are substantial problems for $\PTV$. While
the Hartle--Hawking wave function, $\PNB$, has not been found to yield a
definitive prediction of inflation for any potentials, $V(\ph)$, considered to
date \cite{Luk}, the prognosis for $\PTV$ seems to be even more doubtful as an
identification of type (\ref{Iproba}), (\ref{Iprobb}) cannot be made in the
first place if $p\ge1$.

One might wonder whether these conclusions would be altered by the addition of
non-inflationary matter degrees of freedom to the model. Although strictly
speaking the matter degrees of freedom themselves should be quantized, some
indication of their bulk effect might be obtained by adding a perfect fluid
source to the Wheeler-DeWitt equation \cite{Ku,Fi,CCS,RS,CM}. For a
homogeneous isotropic fluid of energy density, $\rh$, and pressure obeying an
equation of state $P=w\rh$, we find that the minisuperspace Wheeler-DeWitt
equation (\ref{Bocher25}) is replaced by
\begin{equation}
z^2{\dd^2y\over\dd z^2}+z{\dd y\over\dd z}+\left(2Vz^3+\be z^{3(1-w)/2}-k z^2-
\nu^2\right)y=0,
\label{WdWgen}\end{equation}
where the spatial curvature, $k$, is identified explicitly for the purpose of
the discussion that follows, $z=\az$ and $\nu=\case14(p-1)$ as before. The
normalization factor in (\ref{action}) is now given in general by $\si^2=\KA/(3
\Vv)$, $\Vv$ being the 3-volume of the unit spatial hypersurface (with suitable
topological identifications for $k=0,-1$), and $\be=2\KA\si^2a\Z0^{3(1+w)}/(3
\rh\Z0)$ for a matter source with energy density normalized to $\rh=\rh\Z0$ at
$a=a\Z0$. For large values of $a$, the scalar potential term dominates
irrespective of the other matter degrees of freedom or the spatial curvature,
and the solutions approximate to a superposition of modes $a^{(1-p)/2}J_{(p-1)/
6}(\case13a^3\sqrt{V})$ and $a^{(1-p)/2}Y_{(p-1)/6}(\case13a^3\sqrt{V})$ as was
discussed in \cite{KW}.

Now let us consider the properties of the solutions as $a\rarr0$, which is
important for normalization of the wave function. The case of stiff matter $(w=1
)$ in general yields a singular wave function as $a\rarr0$ for {\it both} modes,
and hence any choice of boundary condition, the singularity being of infinitely
oscillatory type if $\be>\nu^2=\case1{16}(p-1)^2 $,
\begin{equation}
\PS\;\goesas_{a\rarr0}\;a^{(1-p)/2}\left\{C\Z1\cos\left[\ln(\case12\sqrt{\be-
\nu^2}\,a^2)\right]+C\Z2\sin\left[\ln(\case12\sqrt{\be-\nu^2}\,a^2)\right]
\right\},
\end{equation}
where $C\Z{1,2}$ are constants.
This is a pathological type of behaviour which does not distinguish between
different boundary conditions for the wave function, and is presumably not of
physical interest. Apart from this case, however, one may readily see that
the conclusions of \cite{KW} are not altered in general by the presence of
additional matter, since the indicial equation for series solutions about
$a=0$, which primarily determines the properties of the solutions near $a=0$,
is unaffected for $w<1$.

Consider the case of radiation ($w=\case13$), for example, in which case
(\ref{WdWgen}) is still of B\^ocher type $\{25\}$ \cite{MS}. As $a\rarr0$ the
leading order behaviour is given by that of the exact solutions to
(\ref{WdWgen}) with $V=0$, which are given by \cite{RS}
\begin{eqnarray}
\PS&=&a^{-(2\nu+1)}\left\{C\Z1 M_{\be/2,\,\nu}(a^2)+
C\Z2W_{\be/2,\,\nu}(a^2)\right\},\nonumber\\
&=&\exp[-\az]\left\{C\Z1\Mfn{\nu+\case12-\case\be2,\,1+2\nu,\,a^2}
+C\Z2\Ufn{\nu+\case12-\case\be2,\,1+2\nu,\,a^2}\right\},
\label{whit1}\end{eqnarray}
in terms of Whittaker functions and Kummer functions \cite{AbS} respectively,
for $k=+1$. Since $\exp[-\az]\Mfn{\nu+\case12,\,1+2\nu,\,a^2}=\GA[\nu+
1]\left(2\over a\right)^{2\nu}I_\nu\Az$ and $\exp[-\az]\Ufn{\nu+\case12,\,1+
2\nu,\,a^2}=\pi^{-1/2}a^{-2\nu}K_\nu\Az$ \cite{AbS}, we of course
retrieve the limiting $V=0$ solutions discussed in previous sections when
$\be=0$. The solutions for $k=-1$ and $V=0$ are likewise given by
\begin{eqnarray}
\PS=&&\case12C\Z1\left\{\e^{ia^2/2}\Mfn{\nu+\case12-i\case\be2,1+2\nu,-ia^2}+
\e^{-ia^2/2}\Mfn{\nu+\case12+i\case\be2,1+2\nu,ia^2}\right\}\nonumber\\ &+&
\case12C\Z2\left\{\e^{-i\pi\nu+ia^2/2}\Ufn{\nu+\case12-i\case\be2,1+2\nu,
-ia^2}+\e^{i\pi\nu-ia^2/2}\Ufn{\nu+\case12+i\case\be2,1+2\nu,ia^2}\right\}
\nonumber\\
\label{whit2}\end{eqnarray}
which reduces to a superposition of ordinary Bessel function modes
\begin{equation}
\PS=a^{(1-p)/2}\left\{B\Z1J_\nu\Az+B\Z2Y_\nu\Az\right\},
\end{equation}
when $\be=0$, where relative to (\ref{whit2}) $B\Z1=2^{2\nu}\GA[\nu+1]C\Z1$
and $B\Z2=-\case12\sqrt{\pi}C\Z2$.
In the spatially flat case, $k=0$, the solution to (\ref{WdWgen}) with $V=0$
is given in terms of ordinary Bessel functions modes for all values of $\be$
\cite{CCS,CM}
\begin{equation}
\PS=a^{(1-p)/2}\left\{C\Z1J_{2\nu}(\sqrt{\be}\,a)+C\Z2Y_{2\nu}(\sqrt{\be}\,a)
\right\}.\label{whit3}
\end{equation}

In each of the cases (\ref{whit1}), (\ref{whit2}) and (\ref{whit3}), the
mode with coefficient $C\Z1$ is finite as $a\rarr0$, whereas the mode with
coefficient $C\Z2$ is proportional to $a^{-4\nu}=a^{1-p}$ as $a\rarr0$, which
causes $\PS$ to diverge for $p\ge1$ unless $C\Z2=0$, in direct analogy to the
radiation-free solutions ($\be=0$). Unfortunately, the requirement that
Vilenkin's wave function, $\PTV$, is outgoing means that the coefficients of
{\it both} modes must be non-zero, and $C\Z2=0$ is not possible for $\PTV$.
In particular, in the model here Vilenkin's boundary condition requires
that
\begin{equation}
\w{Re} {i\over\PTV}{\pt\PTV\over\pt a} > 0,
\end{equation}
in the Lorentzian region, which includes all values $a>0$ if $k=0,-1$. In the
$k=0$ case (\ref{whit3}), for example, the linear combination of modes which
satisfies this condition is
\begin{equation}
\PTV\propto a^{(1-p)/2}\left\{J_{2\nu}(\sqrt{\be}\,a)-iY_{2\nu}(\sqrt{\be}\,a)
\right\},
\end{equation}
namely a Hankel function of the second kind, for which $\w{Re} {i\over\PTV}
{\pt\PTV\over\pt a}=2[\pi a({J_{2\nu}}^2+{Y_{2\nu}}^2)]^{-1}$. Thus the
problems faced by Vilenkin's wave function, especially with regard to the
prediction of inflation \cite{KW}, appear to be generic.

One must also conclude that great care must be exercised when making arguments
based on the identification of probabilities according to (\ref{Iproba}),
(\ref{Iprobb}). In general, the contribution of the prefactor and choices which
depend on Planck scale physics cannot be ignored when one is considering the
problem of the nucleation of the universe.

\bigskip\noindent{\bf Acknowledgement} I would like to thank the Australian
Research Council for financial support.
\global\preprintstytrue \def\refname{\protect\small{\bf REFERENCES}}

\end{document}